\documentclass[10pt,journal,compsoc]{IEEEtran}
\ifCLASSOPTIONcompsoc
  \usepackage[nocompress]{cite}
\else
  \usepackage{cite}
\fi
\ifCLASSINFOpdf
  \usepackage[pdftex]{graphicx}
\else
\fi
\usepackage{amsmath,amssymb,amsfonts}
\usepackage[linesnumbered, ruled,vlined]{algorithm2e} 
\usepackage{xcolor}
\usepackage{algpseudocode} 
\usepackage{graphicx}
\usepackage{textcomp}
\def\BibTeX{{\rm B\kern-.05em{\sc i\kern-.025em b}\kern-.08em
    T\kern-.1667em\lower.7ex\hbox{E}\kern-.125emX}}

\usepackage{textcomp}
\usepackage{caption}
\usepackage{subcaption}
\usepackage{amscd}
\usepackage{amsthm,enumerate}

\usepackage{pgfplots}
\usepackage{pgfplotstable}
\pgfplotsset{compat=1.7}
\usepackage{tikz}

\begin{document}
%
\title{A Tractable Probabilistic Approach to Analyze Sybil Attacks in Sharding-Based Blockchain Protocols}
%
%
%
%

\author{Abdelatif~Hafid,~\IEEEmembership{Member,~IEEE,}
        Abdelhakim~Senhaji~Hafid,~\IEEEmembership{}
        and~Mustapha~Samih,~\IEEEmembership{}
\IEEEcompsocitemizethanks{\IEEEcompsocthanksitem A. Hafid and A. S. Hafid are with Montreal Blockchain Laboratory (mbl), Department of Computer Science and Operational Research, University of Montreal, Montreal, QC H3T 1J4, Canada.\protect\\
E-mail: abdelatif.hafid@umontreal.ca
\IEEEcompsocthanksitem A. Hafid and M. Samih are with Team of EDA – Mathematical Laboratory and their Applications, Department of Mathematics, Faculty of Sciences, University of Moulay Ismail, Meknes 50050, Morocco.}

\thanks{This work was supported in part by the Mohammed VI Polytechnic University - UM6P.}} 

%
%

\markboth{}%
{Hafid \MakeLowercase{\textit{et al.}}: A Tractable Probabilistic Approach to Analyze the Security of Sharding-Based Blockchain Protocols}
%



\IEEEtitleabstractindextext{%
\begin{abstract}
Blockchain like Bitcoin and Ethereum suffer from scalability issues. Sharding is one of the most promising and leading solutions to scale blockchain. The basic idea behind sharding is to divide the blockchain network into multiple committees, where each processing a separate set of transactions, rather than the entire network processes all transactions. In this paper, we propose a probabilistic approach to analyze the security of sharding-based blockchain protocols. Based on this approach, we investigate the threat of Sybil attacks in these protocols. The key contribution of our paper is a tractable probabilistic approach to accurately compute the failure probability that at least one committee fails and ultimately compute the probability of a successful attack. To show the effectiveness of our approach, we conduct a numerical and comparative analysis of the proposed approach with existing approaches.
\end{abstract}

}

\maketitle

\IEEEdisplaynontitleabstractindextext

%
\IEEEpeerreviewmaketitle

\IEEEraisesectionheading{\section{Introduction}\label{sec:introduction}}

%
%
%
%

\IEEEPARstart{S}{ince} the inception of Bitcoin \cite{bitcoin}, interest in blockchain technology, from Industry and Academia, has been booming. Moreover, Blockchain has been used extensively in almost all industry segments, including  internet of things \cite{IoT1, IoTsecurity, IoT3}, cryptocurrencies \cite{bitcoin, ethereum}, education \cite{education}, the healthcare sector \cite{HC1}, the industrial sector \cite{industrial}, and the financial Sector \cite{kirti}. The inherent characteristics of blockchain provide properties like transparency, decentralization, auditability, and security. A blockchain is a distributed database that is organised as a list of ordered blocks, where the committed blocks are immutable. With all these attractive characteristics, one of the key limitation of blockchain is \textit{scalability} \cite{hafid2020scaling}; indeed, the number of transactions that can be processed per second is small (e.g., up to 7 for Bitcoin \cite{bitcoin} and 15 for Ethereum \cite{ethereum}). This is unacceptable for most traditional centralized payment systems that require 1000s of transactions per second (tx/s) (e.g., Visa handles an average of 1700 tx/s \cite{visa}). A number of solutions to scale blockchain have been proposed; we can classify them into two categories: $(1)$ On-chain solutions: they propose modifications to the blockchain protocols, such as sharding (e.g., \cite{rapidchain}, \cite{ethereum}) and block size increase (e.g., \cite{garzik2015block}); and $(2)$ off-chain solutions (aka layer $2$ solutions): these are built on the blockchain protocols; they process certain transactions (e.g., micro-payment transactions) outside the blockchain and only record important transactions (e.g., final balances) on the blockchain. Examples of layer $2$ solutions include Lightning Network \cite{poon2016bitcoin}, Raiden Network \cite{network2018cheap}, and Plasma \cite{poon2017plasma}. 

In this paper, we focus on the sharding solution. The key idea behind sharding is to divide or split the network into subsets, called shards/committees; throughout the paper, we will use the terms shard and committee interchangeably. Each shard will be working on different set of transactions rather than the entire network processing the same transactions. This idea allows the network to scale with the number of shards. It achieves high efficiency for both throughput and storage but may compromise security. Indeed, for the blockchain to be secure, all shards need to satisfy the byzantine validator limit (aka, committee resiliency, i.e., maximum percentage of malicious nodes that a committee can support). For RapidChain \cite{rapidchain}, this limit is $\frac{1}{2}$ (50\%).

\begin{figure}[ht]
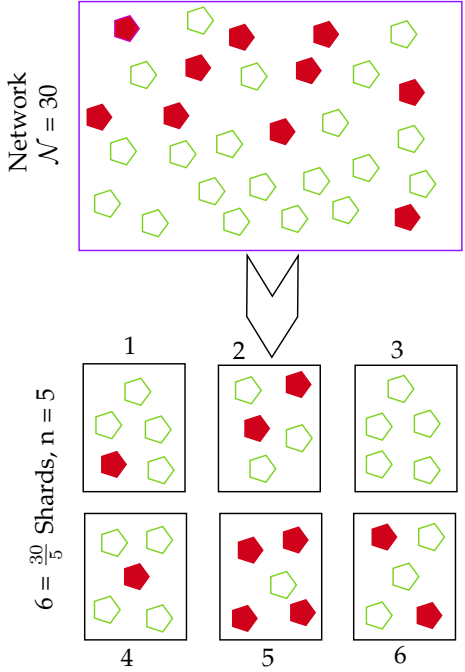

\centering
\tikzset{every picture/.style={line width=0.55pt}} 


\caption{Sharding divides the network into subsets (shards), which means only a shard can handles a set of transactions rather than the entire network. A scenario where there is a single shard takeover attack (shard 5 in this case).}
\label{fig:1}
\end{figure}

The key challenge is that even if the network satisfies the limit, a shard could be compromised. For example, for RapidChain \cite{rapidchain}, let us assume that the network consists of $6$ shards; each shard contains $5$ nodes with 33\% of nodes are malicious (i.e., $10$ nodes are malicious) where $1$ shard has $4$ of the malicious nodes and the rest evenly distributed in the other $5$ shards. In this case, one shard will be $80\%$ byzantine/malicious and thus the entire network will be insecure (because the maximum number of malicious nodes that a shard can support is $50\%$). This is known as a \textbf{\textit{single shard takeover attack}}, see Figure \ref{fig:1}.

\begin{figure*}[ht]
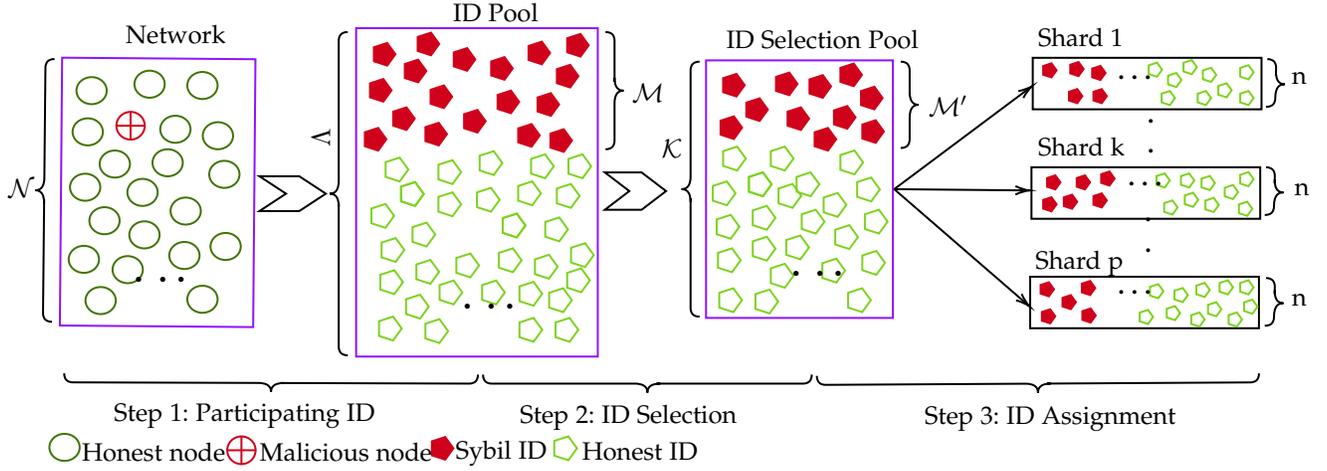

\centering
\tikzset{every picture/.style={line width=0.75pt}} 

%
\caption{Sharding-based blockchain protocol model. Step 1: Each node participates by its ID, save a malicious one, which can participate with more than one. Step 2: Using a consensus mechanism (e.g. PoW), each node competes to add its corresponding ID/IDs to the ID Selection Pool. Step 3: Random distribution of IDs from the ID Selection Pool to shards uniformly.} 
\label{fig:2}
\end{figure*}

In particular, we analyze the security of sharding-based blockchain protocols by computing the failure probability using generating function. Based on this approach, we investigate Sybil attacks and identify the parameters that can deter such attacks. More specifically, we measure the security of sharding protocols by counting the number of years to fail taking into account the failure probability of each shard (i.e., the probability that a shard exceeds the committee resiliency). To do this, we make use of generating functions. These functions are widely used in computer sciences. Rajan et al. \cite{rajan} described many applications of generating functions in engineering and applied sciences. Kohei et al. \cite{akutsu2019analysis} used joint generating function to analyze the retrial queues for cognitive wireless networks with sensing time. In blockchain systems, Wenjuan et al. \cite{zhao2019analysis} used generating function to analyze the average confirmation time of transactions. Jelena et al. \cite{misic2019modeling} used probability generating function to characterize the distribution of the number of connections per node in Bitcoin network; the objective was to model the Bitcoin delivery network.

The key contribution of this paper is to propose a tractable approach to calculate the probability that at least one committee fails using generating function and investigate Sybil attacks based on these probabilities. Note that in sharding-based blockchain protocols, the network is compromised if only one shard is compromised (i.e., 1\% attack); this is why we compute the probability that at least one committee fails. The proposed approach outperforms, in terms of the computation accuracy, the mathematical models proposed in existing contributions \cite{hafid2019methodology}, \cite{hafid2019model}, \cite{rajab2020feasibility}; it also achieves similar computation accuracy as JHDA  \cite{hafid2020joint} in estimating the failure probability, but with much less computational complexity.

The limitations in \cite{hafid2019methodology}, \cite{hafid2019model}, \cite{rapidchain}, \cite{rajab2020feasibility} come from the fact that they assume that the failure probability in the first committee is indicative of the failure probability in any other committee; more specifically, they assume that the failure probability of one epoch is the failure probability of the first committee times the number of committees \cite{hafid2019methodology}, \cite{hafid2019model}, \cite{rapidchain}. However, when the sampling is done without replacement, the samples are not independent; this means that when we sample the first committee, it is clear that the parametrizations of the model change (i.e., the number of nodes in the network which is the number of IDs, as well as the number of malicious nodes which is the number of Sybil IDs. Thus, the failure probability of the second committee will be different from the first, and the third will be different from the first and the second, until the last committee. In addition, there are significant changes in the various parameters from the first to the second to the last sampled committee; this means that the inaccuracy of the estimate proposed in \cite{hafid2019methodology}, \cite{hafid2019model}, \cite{rajab2020feasibility}, grows with the number of committees. In a more recent work, Hafid et al. \cite{hafid2020joint} filled this gap by proposing an approach, called JHDA, that takes into consideration the failure probability of each shard. However, the key limitation of this approach (i.e., JHDA) is its complexity; indeed, it is not practical to accurately compute the failure probability using JHDA; we can only estimate it by executing a large number of trials \cite{hafid2020joint}. In this paper, we address this gap by proposing a new approach called Probabilistic Generating Function Approach (PGFA). Furthermore, we investigate the threat of Sybil attacks based on the probabilities of failure computed by PGFA. Specifically, the contributions of this paper can be summarized as follows:
\begin{itemize}
    \item We propose a tractable probabilistic methodology to analyze the security of sharding-based blockchain protocols using generating function;
    \item We investigate the threat of Sybil attacks based on these probabilities;
    \item We compare, in terms of the computational complexity, PGFA with the the most accurate existing approach (i.e., JHDA); 
    \item We identify the parameters that can reduce the impact (in terms of threat severity) of Sybil attacks.
\end{itemize}

The paper is organized as follows. Section \ref{sec: Analytical Model} presents the proposed Probabilistic Generating Function Approach (PGFA) and compares it (in terms of complexity) with JHDA. Section \ref{sec: Results and Evaluation} evaluates PGFA and compares it with Break Consensus Protocol (BCP) \cite{rajab2020feasibility}. Finally, Section \ref{sec: Conclusion} concludes the paper.

\section{Analytical Model} \label{sec: Analytical Model}
In this section, we describe our approach, PGFA, to calculate/compute the probability that at least one committee fails using generating function.
\subsection{Abbreviations and Definitions}\label{AA}
Table $\ref{table:1}$ shows the list of symbols and variables that are used to describe the proposed PGFA as well as JHDA. 
\begin{table}[ht]
\caption{Notations}
\label{table:1}
\setlength{\tabcolsep}{3pt}
\begin{tabular}{|p{50pt}|p{180pt}|}
\hline
Notation& 
Description \\
\hline
$\mathcal{N}$  & 
Total number of nodes \\
$\Lambda$  & 
Total number of IDs in the ID Pool ($\Lambda = \mathcal{N} - 1 + \mathcal{M}$) \\
$\mathcal{M}$  & 
Total number of Sybil IDs in the ID Pool \\
$\mathcal{M'}$  & 
Total number of Sybil IDs in the ID Selection Pool ($ \mathcal{M'} \leq \mathcal{M} $) \\
$m_{i}$ & 
Number of Sybil IDs in shard i \\
$\mathcal{K}$  & 
Total number of IDs in the ID selection Pool ($ \mathcal{K} \leq \Lambda $)\\
$n$             &
Size of the committee\\
$r$             &
Committee resiliency\\
$\mathcal{R}$    &
Resiliency of the ID Selection Pool\\
$\lambda$         &
Number of committees\\
$p_{e}$ &
Epoch failure probability\\
$\mathcal{A}$&
Average number of years to failure\\
$E_{s}$ &
Expected number of sharding rounds until failure \\
$N_{s}$&
Number of sharding rounds per year\\
$t$&
Number of trials\\
$\left[ x^{\kappa} \right] \Psi(x)$ &
Coefficient of $x^{\kappa}$ in $\Psi(x)$ \\
$\mathcal{P}$&
Probability of selecting at least $m$ Sybil IDs from the ID Pool \\
$\mathcal{P^{'}}$&
Probability that at least one committee fails\\
$\mathcal{P^{''}}$ &
Probability of a successful attack\\
\hline
\end{tabular}
\end{table}

\textbf{Definition 1}. Ordinary Generating Function. The ordinary generating function for the sequence $(u_{z})_{z \geq 0}$ is the power series, which can be expressed as follows:
 \begin{equation}
u_{0} + u_{1}x^{1} + u_{2}x^{2} + u_{3}x^{3} +  \cdots   \label{eq:1}
\end{equation}

\textbf{Definition 2}. Committee Resiliency. The maximum percentage of Sybil IDs that the committee can contain/support whereas still being secure. 

\textbf{Definition 3}. Total Resiliency. The maximum percentage of malicious nodes that the whole network can contain whereas still being secure. It is the maximum percentage of malicious nodes that the entire network can tolerate/support without compromising its security.

\textbf{Definition 4}. ID Pool. ID Pool contains IDs (Sybil and honest IDs) generated by the nodes (honest or malicious nodes); each node generates its own ID, save the ones, which can generate many Sybil IDs.  

\textbf{Definition 5}. Adversary's hash-rate. The percentage of adversary's hash-power in participating in the ID Selection Pool (e.g., if an adversary controls $30$ Sybil IDs in an ID Selection Pool that contains 100 IDs, thus the adversary's hash-rate will be $30\%$).

\textbf{Definition 6}. ID Selection Pool. ID Selection Pool is constructed in each epoch using a consensus mechanism (e.g. \textit{PoW}) and contains the total number of required and valid/qualified IDs; it contains the number of IDs that is necessary in participating in this epoch; this number depends of the security of the network. 

\textbf{Remark}. It is worth noting that a honest node can participate with many honest IDs.

\subsection{Probability of Selecting Sybil IDs from the ID Pool}
In this section, we compute the probability of selecting Sybil IDs from the ID Pool during the construction of the ID Selection Pool.

Figure \ref{fig:2} shows a worst-case scenario of a sharding-based blockchain protocol. In this scenario, we assume that each honest node is able to generate its own ID and one strong (in terms of hash rate) malicious node (aka one adversary) generates numerous Sybil IDs. The sharding process consists of $3$ steps. In the first step, each node participates by its own ID, save a malicious node, it participates with $\mathcal{M}$ IDs. In the second, we construct the ID Selection Pool from the ID Pool. More specifically, using a consensus mechanism (e.g. PoW and PoS), each node competes to generate a valid/qualified ID to join the ID Selection Pool, save the malicious, it adds $\mathcal{M^{'}}$ Sybil IDs ($\mathcal{M^{'}} \leq \mathcal{M}$) thanks to its high hash power. 
In the third step, we randomly distribute (i.e., random assignment) IDs from from the ID Selection Pool to shards.




A few sharding-based Blockchain protocols (e.g. Ethereum sharding \cite{ethereumsharding}) adopt the PoS-based node/ID selection method to select shard members to defend against Sybil attacks. However, in most sharding-based blockchain protocols (e.g., Elastico \cite{elastico}, OmniLedger \cite{omniledger} and RapidChain \cite{rapidchain}), \textit{Proof-of-Work} (\textit{PoW}) consensus is typically used to establish the committee/shard members (committee formation) and generate the corresponding IDs; Byzantine Fault Tolerant (BFT) is used for the intra-committee consensus, which is used within a committee to create and append the blocks \cite{rapidchain}, \cite{hafid2020scaling}. More specifically, nodes who want to join/stay in the protocol use \textit{PoW} consensus to generate valid IDs (i.e., public keys). Only the nodes that can solve the ID generation \textit{PoW} puzzle (e.g. RapidChain uses a fresh fixed \textit{PoW} puzzle \cite{rapidchain}) can generate valid IDs. It turns out that the nodes that have a higher hash-rate have a higher probability to solve the ID generation \textit{PoW} puzzle compared to those of lower hash-rate.
The security of the network will not be compromised as long as the malicious node's computational power is limited.

Now, let $\mathcal{X}$ be a random variable that counts the number of Sybil IDs selected from the ID Pool. In the following, we compute the probability of selecting \textit{exactly} $m$ Sybil IDs from the ID pool (\textbf{Lemma 1}) and then the probability of selecting \textit{at least} $m$ Sybil IDs from the ID pool (\textbf{Lemma 2}). 

\textbf{Lemma 1:} In a sharding-based blockchain protocol, the probability of selecting \textit{exactly} $m$ Sybil IDs from the ID pool is given by: 
\begin{equation} 
    P(\mathcal{X} = m) = \frac{ \binom{\mathcal{M}}{m}  \binom{\mathcal{N} -1}{\mathcal{K} -m} }{ \binom{\Lambda}{\mathcal{K}} }
\end{equation}

\textbf{Proof:} In a sharding-based blockchain protocol, the process of assigning IDs to shards can be modeled as a random sampling without replacement, because the shards can not overlap. In this case, the hypergeometric distribution yields a better probability approximation compared to any other probability distribution, including that of Binomial's \cite{distinguishing}. Thus, the proof of \textbf{Lemma 1} is a direct result from the fact that $X$ follows a hypergeometric distribution \cite{distinguishing}, \cite{hafid2020joint}.   

\textbf{Lemma 2:} In a sharding-based blockchain protocol, the probability of selecting \textit{at least} $m$ Sybil IDs from the ID pool can be computed as follows: 
\begin{equation} 
    P(\mathcal{X} \geq m ) = \sum_{s = m}^{\mathcal{K}} \frac{ \binom{\mathcal{M}}{s}  \binom{\mathcal{N} -1}{\mathcal{K} -s} }{ \binom{\Lambda}{\mathcal{K}} }
\end{equation}

\textbf{Proof:} Given the fact that $\mathcal{X}$ follows a hypergeometric distribution, the proof of \textbf{Lemma 2} can be extracted directly from the cumulative hypergeometric distribution \cite{distinguishing}.

The following subsection will be devoted to compute and calculate the probability that at least one shard fails. 
\subsection{Probability that at least one Shard Fails}
In this section, we compute the probability that at least one shard fails in a sharding-based blockchain protocol using the proposed approach (i.e., PGFA). First, we briefly describe how JHDA \cite{hafid2020joint} computes such a probability. Then, we present the details of the proposed PGFA. 
\subsubsection{Joint Hypergeometric Distribution Approach (JHDA)} 
In a more recent work, Hafid et al. \cite{hafid2020joint} proposed a novel methodology-based joint hypergeometric distribution to analyze the security of sharded protocols, which can be summarized in \textbf{Theorem 1} (see proof in \cite{hafid2020joint}).

Now, let $\mathcal{Y}_i$ be a random variable that computes the number of Sybil IDs in shard $i$. \textbf{Theorem 1} computes the probability that at least one shard fails using JHDA.

\textbf{Theorem 1:} In a sharding-based blockchain protocol, the probability that at least one shard fails using JHDA can be expressed as follows:
\begin{equation} \label{eq:11}
 \mathcal{P^{'}} = 1 - P(\mathcal{Y}_i \leq nr, i \in \{1, 2, \dots, \lambda \} \end{equation}
where 
\begin{eqnarray}
 P(\mathcal{Y}_i \leq nr, i \in \{1, 2, \dots, \lambda \}) \\
= \sum_{m_1=0}^{nr} \sum_{m_2=0}^{nr} \dots \sum_{m_\lambda =0}^{nr} \prod_{i=1}^{\lambda} \binom{n}{m_i} \bigg / \binom{\mathcal{K}}{\mathcal{M'}} 
\end{eqnarray} 
The limitation of JHDA comes from the fact that this approach requires a high computation power due to its high complexity. We address this limitation in the following subsection, where we introduce a Probabilistic Generating Function Approach (PGFA) as an alternative solution to analyze the security of sharding-based blockchain protocols.   
\subsubsection{Generating Function Approach}
Generating function transforms problems about sequences into problems about functions. With generating function, we can then apply many and several machinery problems to problems about sequences. They can also be used to find closed-form expressions for sums, to solve counting problems, and to solve recurrence relations \cite{lehman2010mathematics}. There are few classes of generating function in common use (e.g., exponential generating function \cite{petersen2019inquiry} and Ordinary Generating Function (OGF) \cite{lehman2010mathematics}). In this paper, we use the ordinary class, because it is useful for solving counting problems; in particular, problems involving choosing items (i.e., sampling with/without replacement) from a set (entire network in our case). Ordinary generating function does often lead to a polynomial function by letting the coefficient of $x^{k}$ be the number of ways to choose $k$ items (nodes in our case).

\subsection{Modeling}
In sharding-based blockchain protocols, the process of assigning IDs to committees (or partition of the network into committees/shards) can be defined as a sampling without replacement, because the committees do not overlap \cite{hafid2019model}. When the sample is done without replacement, we make use of the hypergeometric distribution instead of binomial's \cite{hafid2019model}, \cite{distinguishing}. Note that to model this sampling using the hypergeometric distribution, we need to make use of joint hypergeometric distribution to cover the failure probabilities of all committees, which is a complex/difficult to compute; it is a closed-form expression. This is the reason why we choose generating function. Generating function is fundamentally devoted to such complicated problems.

The generating function corresponding to select \textbf{\textit{distinct}} items from a finite set (as sampling \textbf{\textit{without replacement}}) can be expressed as follows:  
\begin{equation}
\binom{n}{0} + \binom{n}{1}x^{1} + \binom{n}{2}x^{2} + \binom{n}{3}x^{3}+ \dots + \binom{n}{k} x^{k }  \label{eq:sampling_without_replacement}
\end{equation} 
where \(\binom{n}{k}\) is the number of ways/possibilities to choose/select $k$ distinct items from a set of size $n$; it can be expressed as follows:
\begin{equation}
\binom{n}{k} = \frac{n!}{k! (n - k)!} \label{eq:binom}
\end{equation}
The main aim of our analysis is how to divide $m$ malicious nodes between $\lambda$ committees without exceeding the resiliency of each committee. Let $ m = m_{1} + m_{2} + \dots + m_{\lambda} $ where $m_{1}$ is the number of Sybil IDs in the first committee, $m_{2}$ is the number of Sybil IDs in the second committee, and so on. We need to divide $m$ Sybil IDs, whereas for all $i \in \{ 1, 2, \cdots, \lambda \}$,  $m_{i}$ is smaller than the resiliency of the committee (e.g., $r = \frac{1}{2}$ for RapidChain \cite{rapidchain}). 

Now, based on \eqref{eq:sampling_without_replacement}, the generating function that represents one committee can be expressed as follows:
\begin{equation}
\psi(x) = \binom{n}{0} + \binom{n}{1}x^{1} + \binom{n}{2}x^{2} + \dots + \binom{n}{\lfloor nr \rfloor} x^{\lfloor nr \rfloor }  \label{eq:shard}
\end{equation}
Precisely, if we assume \eqref{eq:shard} refers to shard 1, thus $m_{1}$ can take 0 , 1 , 2 , \dots, or  $\lfloor nr \rfloor$ Sybil IDs. And if \eqref{eq:shard} refers to shard 2, $m_{2}$ also can take 0 , 1 , 2 , \dots, or  $\lfloor nr \rfloor$ malicious nodes, and so on for the other shards. Based on \eqref{eq:shard}, the number of possibilities for shard 1 to receive $m_{1}$ = 0 malicious nodes (i.e., shard 1 does not contain any malicious node) is $\binom{n}{0}$, the number of possibilities for shard 1 to receive $m_{1}$ = 1 malicious nodes (i.e., shard 1 contains one malicious node) is $\binom{n}{1}$; thus, the number of possibilities for shard 1 to get $m_{1}$ = $\lfloor nr \rfloor$ Sybil IDs is $\binom{n}{\lfloor nr \rfloor}$. Our first objective is to calculate the number of possibilities we can split $m$ malicious nodes between $\lambda$ committees without exceeding the committee resiliency (i.e., without exceeding the maximum number of Sybil IDs that the committee can tolerate; this number is $\lfloor nr \rfloor$ in \eqref{eq:shard}). Then, we can compute the failure probability, which represents the probability that at least one committee exceeds the committee resiliency.

Generally, the generating function for choosing elements from a union of disjoint sets is the product of the generating functions from choosing from each set \cite{lehman2010mathematics}. In our case, we have $\lambda$ committees which do not overlap (the sampling is done without replacement). Indeed, the entire network is the union of disjoint committees; it can be modeled as follows:
\begin{equation}
C = \bigcup_{i=1}^{\lambda} C_{i} \label{eq:union}
\end{equation} 
where $C$ is a set which contains all nodes in the entire network with  $card(C) = N$, $C_{i}$ contains the number of nodes in committee $i$  with $card(C_{i}) = n $ for all $i \in \{ 1, 2, \cdots, \lambda \}$, and for all $i, j \in \{ 1, 2, \cdots, \lambda \}$, $C_{i} \cap C_{j} = \emptyset $ for $i \neq j$. 
To compute the total number of ways we can distribute the $m$ Sybil IDs across all committees, we multiply this generating function with itself $\lambda$ times:

\begin{equation}
\Psi(x)= \left( \binom{n}{0} + \binom{n}{1}x^{1} + \binom{n}{2}x^{2}+ \dots + \binom{n}{\lfloor nr \rfloor} x^{\lfloor nr \rfloor}\right)^{\lambda}  \label{eq:Psi_1}
\end{equation}
where 
\begin{equation}
\Psi(x)= \left( \psi(x)\right)^{\lambda} \label{eq:Psi_2}
\end{equation}
Now, we need to extract the coefficient of $x^{m}$ in \eqref{eq:Psi_1}, which corresponds to the number of possibilities we can split $m$ Sybil IDs across $\lambda$ committees without exceeding the resiliency of each committee (note that all the committees have the same resiliency). To compute the sequence of coefficients from this generating function, we need to compute the \textit{Taylor series} for this generating function \cite{lehman2010mathematics}. Therefore, the required coefficient can be expressed as follows:  
\begin{equation}
\left[ x^{m} \right] \Psi(x) = \frac{\Psi^{m}(0)}{m!}  \label{eq:coefficient}
\end{equation}
where $ \Psi^{m}(0)$ is the $m$th derivative of $\Psi(x)$ evaluated at $x = 0$.

The required coefficient can be determined explicitly (analytically) by using \eqref{eq:coefficient}; however, this process involves tedious and complex calculations. Instead, we can determine the coefficient computationally by using \textit{SymPy} package \cite{meurer2017sympy}; it uses the symbolic math system making the process easier to execute.

The probability that at least one committee fails is the probability that at least one committee contains more than $nr$ Sybil IDs. This means that the number of Sybil IDs in the committee exceeds the committee resiliency. Therefore, the probability that at least one committee fails using PGFA is expressed in \textbf{Theorem 2}.

\textbf{Theorem 2:} In a sharding-based blockchain protocol, the failure probability that at least one committee fails using PGFA can be expressed as follows:
\begin{equation} 
\mathcal{P^{'}} = 1 - \frac{\left[ x^{m} \right] \Psi (x)}{\binom{\mathcal{N}}{m}}   \label{eq:9}
\end{equation}
where  \(\binom{ \mathcal{N}}{m}\) is the total number of possibilities to select $m$ Sybil IDs from $\mathcal{N}$ IDs.

\subsection{Probability of a Successful Attack}
In this section, we compute the probability of a successful attack (the failure probability of the entire network); this means that we take into consideration the probability of selecting Sybil IDs from the ID Pool as well as the probability of at least one shard takeover attack. 

\textbf{Theorem 3:} Given a sharding-based blockchain protocol, the probability of a successful attack (assumed by an adversary) can be computed as follows:

\begin{equation}
   \mathcal{P^{''}} = \sum_{s = m}^{\mathcal{K}} \frac{ \binom{\mathcal{M}}{s}  \binom{\mathcal{N} -1}{\mathcal{K} -s} }{ \binom{\mathcal{N} -1 + \mathcal{M}}{\mathcal{K}} } \left(  1 - \frac{\left[ x^{m} \right] \Psi (x)}{\binom{ \mathcal{N}}{m}} \right). 
\end{equation}

\textbf{Proof:} By taking into consideration both probabilities (i.e., the probability of selecting Sybil IDs from the ID pool as well as the probability that at least one shard fails), and based on \textbf{Lemma 2} and \textbf{Theorem 2}, the probability of a successful attack ($\mathcal{P^{''}}$) can be expressed as follows.

\begin{equation} \label{eq:successful_attack}
\begin{split}
\mathcal{P^{''}} & = P( \mathcal{X} = nr) \mathcal{P^{'}} + \dots + P(\mathcal{X} = \mathcal{K}) \mathcal{P^{'}} \\
 & = \big ( P(\mathcal{X}= nr) + \dots + P(\mathcal{X} = \mathcal{K}) \big ) \mathcal{P^{'}} \\
 & = \sum_{k=nr}^{\mathcal{K}} P(\mathcal{X} \geq k) \mathcal{P^{'}}
\end{split}
\end{equation}

\subsection{Years to Fail}
To measure the security of a given protocol, we propose to compute the average number of years to failure. To perform this computation, we need to determine the failure probability of epoch per sharding round, which refers to the failure probability that at least one committee fails. The average number of years to fail corresponding to PGFA is given by:
\begin{equation}
\mathcal{A} = \frac{E_{s}}{N_{s}}, \hspace{0.5cm} where \hspace{0.5cm} E_{s} = \frac{1}{\mathcal{P^{''}}}
\label{eq:12}
\end{equation}

The average number of years to fail corresponding to JHDA ($\mathcal{P'}$ must be calculated by \textbf{Theorem 1}) is given by:
\begin{equation}
\mathcal{A} = \frac{E_{s}}{N_{s}}, \hspace{0.5cm} where \hspace{0.5cm} E_{s} = \frac{1}{p_{e}}
\label{eq:13}
\end{equation}
where $$ p_{e} = \mathcal{P} \times \mathcal{P'} $$

\subsection{Complexity Comparison}
Table \ref{table:complexity} shows a comparison (in terms of complexity) between PGFA and JHDA. 
\begin{table}[ht]
\renewcommand{\arraystretch}{1.3}
\caption{Complexity Comparison between PGFA and JHDA.}
\label{table:complexity}
\centering
\begin{tabular}{c||c}
\hline
\bfseries Approach & \bfseries Complexity\\
\hline\hline
JHDA \cite{hafid2020joint} & $O((nr -1)^{\lambda})$\\
PGFA  & O(nr - 1)\\
\hline
\end{tabular}
\end{table}
Specifically, Table \ref{table:complexity} shows that JHDA is very difficult and impractical to compute whereas PGFA shows a linear complexity. Thus, the feasibility of PGFA to analyze the security of sharding-based blockchain protocols.
\section{Results and Evaluation} \label{sec: Results and Evaluation}
In this section, we compare (in terms of accuracy) PGFA and JHDA \cite{hafid2020joint} via simulations. In particular, we compute the probability of selecting Sybil IDs from the ID Pool to construct the ID Selection Pool as well as the failure probability that at least one committee fails using PGFA and JHDA. These probabilities represent the failure probability of the whole network in one sharding round (i.e., one epoch). We also investigate the threat of Sybil attacks and identify the parameters that impact (in terms of threat severity) these attacks. 
\subsection{Simulation Setup}
To implement our approach, we use \textit{SymPy Python library}, which makes the use of symbolic mathematics easier \cite{meurer2017sympy}, (e.g., polynomial functions). Particularly, we use \textit{sympy.Symbol()} and \textit{sympy.Poly()} to declare the variable ``x" explicitly and  to explicitly express the polynomial in \eqref{eq:shard}. To implement JHDA, we refer the readers to \cite{hafid2020joint}.

\begin{algorithm}
\caption{Algorithm for Computing $\mathcal{P^{'}}$ (PGFA)}
\newcommand\mycommfont[1]{\footnotesize\ttfamily\textcolor{blue}{#1}} 
\SetCommentSty{mycommfont} 
\SetKwInput{KwInput}{Input}                
\SetKwInput{KwOutput}{Output}              

\SetKwFunction{FPGFA}{PGFA}  
\SetAlgoLined
\DontPrintSemicolon
\KwInput{$\mathcal{K}$ (IDs in the ID Selection Pool), $n$ (Committee size), $\mathcal{M'}$ (Sybil IDs), $r$ (Committee resiliency) } 
\KwOutput{$\mathcal{P^{'}}$ \tcp*{Probability that at least one shard fails using PGFA} } 
\SetKwProg{Fn}{Function}{:}{}
\Fn{\FPGFA{$\mathcal{K}$, $n$, $\mathcal{M'}$, $r$}}{
    $ \lambda \gets \lfloor \frac{\mathcal{K}}{n} \rfloor$  \tcp*{ Number of shards}
    $a \gets [\hspace{0.2 cm}]$ \tcp*{Empty list}
    \For{$ \forall i \in \mathcal{D} = \langle \lfloor r \times n \rfloor, \lfloor r \times n \rfloor - 1, \dots, 0 \rangle $} 
    {
   $a \gets \langle \binom{n}{ \lfloor r \times n \rfloor}, \binom{n}{\lfloor r   \times n \rfloor- 1}, \dots, 0 \rangle $ \tcc{Vector with $\lfloor r   \times n \rfloor + 1$  components} \;
   $ X \gets \langle x, x, \dots, x \rangle $ \tcc{Vector with $\lfloor r \times n \rfloor + 1$ components} \;
   $ \psi(x) \gets a \cdot X =  \sum_{i= \lfloor r \times n \rfloor}^{0} \binom{n}{i} x $  \tcc{$a \cdot X$ means the dot product of vectors $a$ and $X$. $\psi(x)$ is a polynomial with variable $x$ and coefficients  $\{ \binom{n}{i}, \forall i \in \mathcal{D} \} $ }
   }
   $ \Psi(x) \gets \big ( \psi(x) \big)^{\lambda} = \big( \sum_{i= }^{0} \binom{n}{i} x \big )^{\lambda} $ \tcc{$\psi(x)$ power the number of shards ($\lambda$)}

   $ \zeta \gets [x^{\mathcal{M'}}] \Psi(x) $ \tcc{Coefficient of $x^{\mathcal{M'}}$ in $\Psi(x)$, which corresponds to the number of possibilities to distribute $\mathcal{M'}$ Sybil IDs across $\lambda$ shards (see Equation \ref{eq:coefficient})} 

   $\mathcal{P'} \gets 1 - \frac{\zeta}{\binom{\mathcal{K}}{\mathcal{M'}}}$ \;
     
     \KwRet $\mathcal{P'}$ \;
}
\end{algorithm}

Table \ref{table:2} shows the values of the parameters used in the simulations.
\begin{table}[ht]
\renewcommand{\arraystretch}{1.3} 
\caption{Parameter Settings}
\label{table:2}
\centering
\begin{tabular}{c||c}
\hline
\bfseries Parameter & \bfseries Value\\
\hline\hline
$\mathcal{N}$ & 1000, 1200, 1400\\
$\mathcal{K}$ & 400, 600, 800\\
$\mathcal{M}$ & 200, 250\\
$\mathcal{M'}$ & 125, 200, 250\\
$r$ & 0.25, 0.27, 0.333\\
$\mathcal{R}$ & 0.20, 0.25, 0.333\\
$N_s$ & 365, 185\\
$n$ & 80, 100, 200\\
\hline
\end{tabular}
\end{table}
\subsection{Results and Analysis}
\begin{figure*}[ht]
\centering
\includegraphics[width = 1 \textwidth]{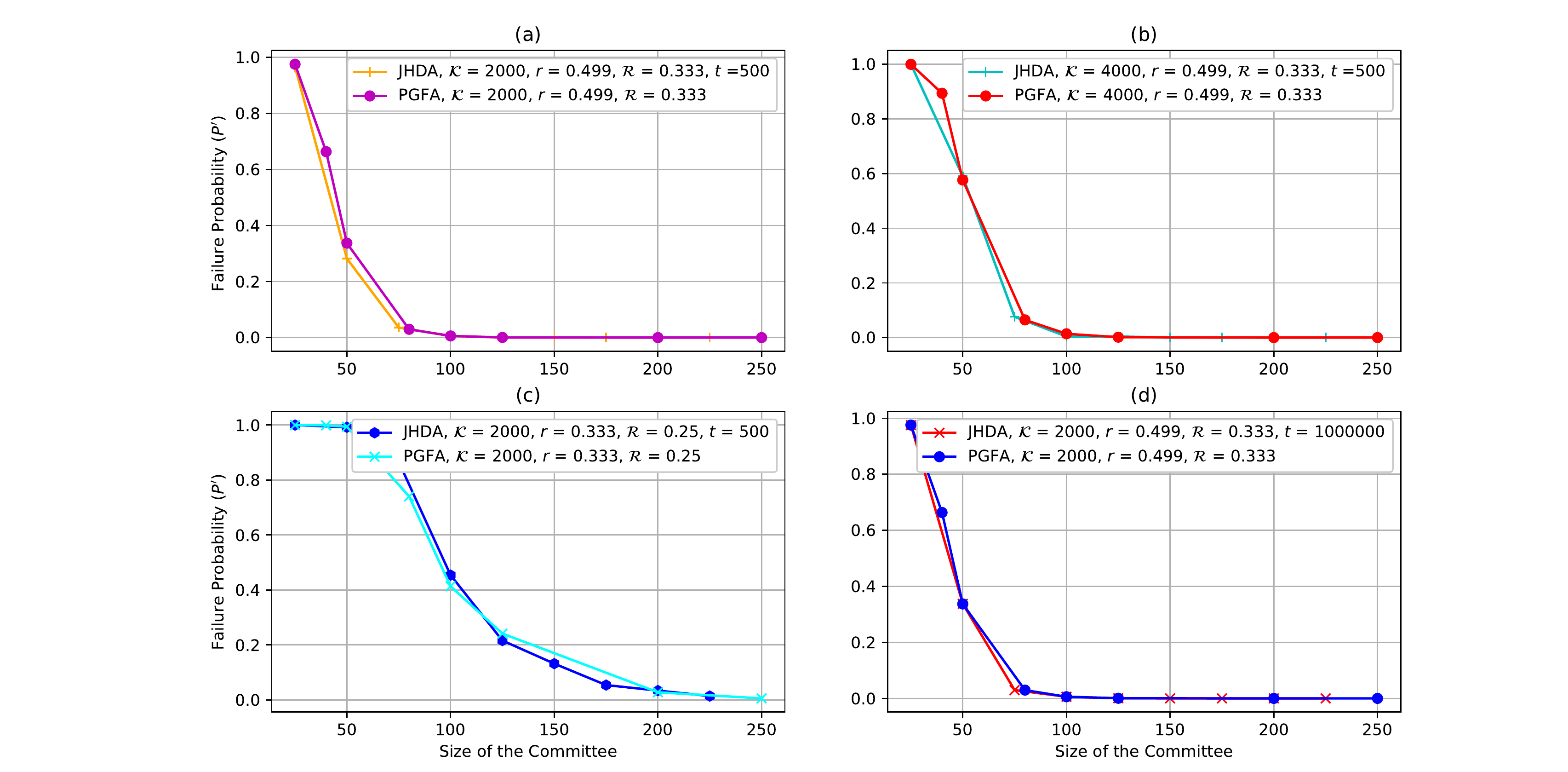}
\caption{Comparison between JHDA and PGFA for different network sizes, values of resiliency, and number of trials.} 
\label{fig:comparison_JHDA_PGFA}
\end{figure*}
Figure \ref{fig:comparison_JHDA_PGFA} shows a comparison between JHDA and PGFA for different ID Selection Pool sizes ($\mathcal{K}$), values of resiliency ($r$ and $\mathcal{R}$), and numbers of trials ($t$) when varying the size of the committee from 25 to 250 by a step of 25.
 
More specifically, Figures \ref{fig:comparison_JHDA_PGFA}(a), \ref{fig:comparison_JHDA_PGFA}(b), and \ref{fig:comparison_JHDA_PGFA}(c) show that the performance of FPGA is close to JHDA. Figure \ref{fig:comparison_JHDA_PGFA}(d) shows a much closer match between the results achieved by PGFA and JHDA. This is expected since, for JHDA, when the number of trials ($t$) increases the estimated failure probability moves towards the exact failure probability \cite{hafid2020joint}. We conclude that the proposed PGFA allows for an accurate computation of the failure probability.

\begin{figure*}[ht]
\centering
\includegraphics[width= 1 \textwidth]{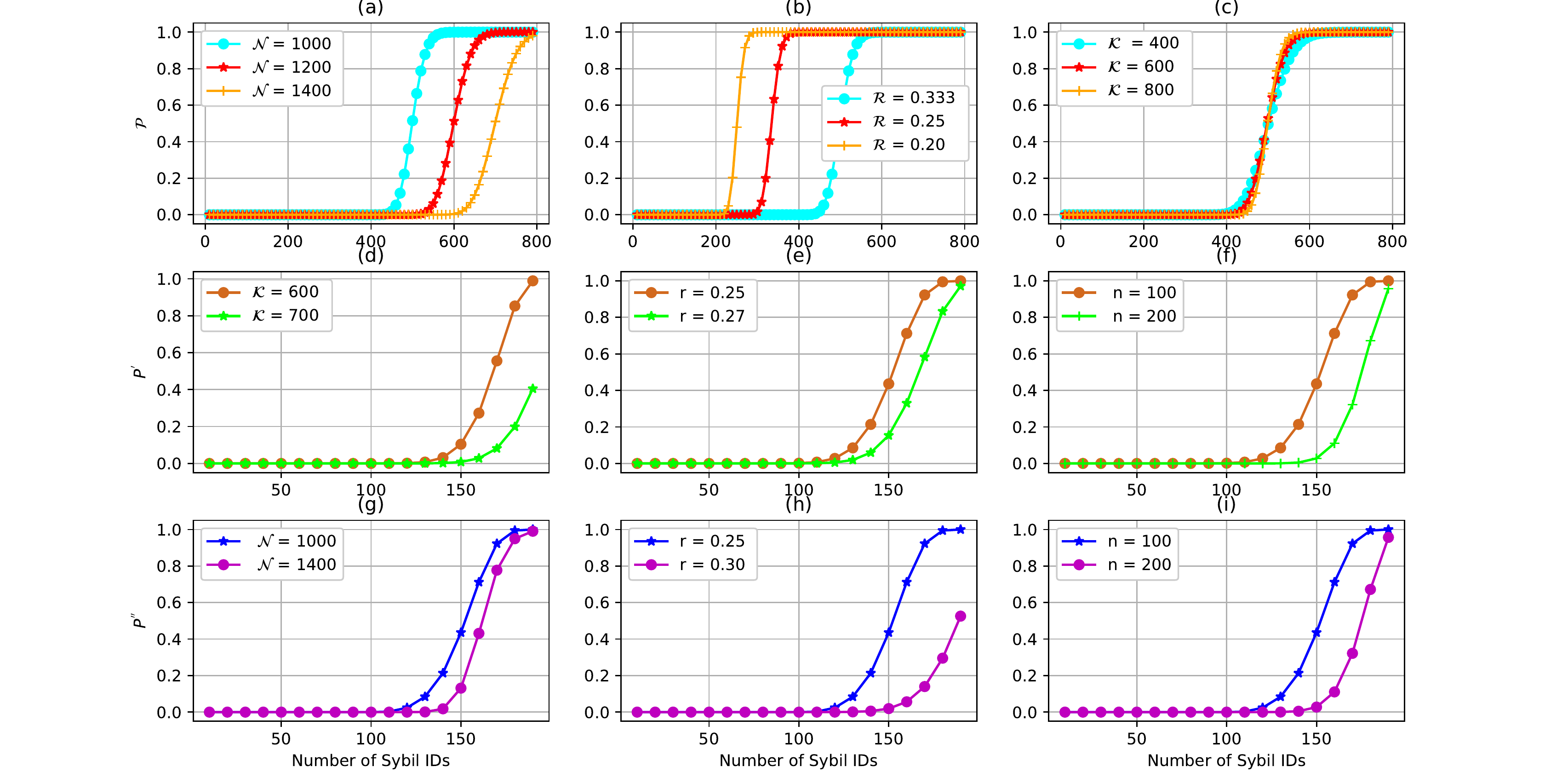}
\caption{ Performance of PGFA on computing failure probability of selecting Sybil IDs from the ID Pool ($\mathcal{P}$), failure probability for at least one shard takeover attack ($\mathcal{P'}$), and probability of a successful attack ($\mathcal{P''}$) when varying the number of Sybil IDs from 10 to 200 (by a step of 10 IDs) for different network sizes ($\mathcal{N}$), values of committee resiliency ($r$), values of ID Selection Pool resiliency ($\mathcal{R}$), ID Selection Pool sizes ($\mathcal{K}$), and for different values of the size of the committee (n).}
\label{fig:r}
\end{figure*}

Figure \ref{fig:r} shows the failure probability of selecting Sybil IDs from the ID Pool ($\mathcal{P}$), the failure probability for at least one shard takeover attack ($\mathcal{P'}$), and the probability of a successful attack ($\mathcal{P''}$) when varying the number of Sybil IDs (we assume that $\mathcal{M}$ = $\mathcal{M'}$; the worst case that can happen) from 10 to 200 (by a step of 10 IDs) for different network sizes, values of committee resiliency, values of ID Selection Pool resiliency, ID Selection Pool sizes, and for different values of committee size (n). These results show that the failure probability increases with the number of Sybil IDs means a high hash-rate of the adversary; this increases the probability to compromise the security of the network. In particular, Figures \ref{fig:r} (a), \ref{fig:r} (b), and \ref{fig:r} (c) show the failure probability of selecting Sybil IDs from the ID Pool for different network size ($\mathcal{N}$ = 1000, $\mathcal{N}$ = 1200, and $\mathcal{N}$ = 1400), ID Selection Pool resiliency ($\mathcal{R}$ = 0.333, $\mathcal{R}$ = 0.25, and $\mathcal{R}$ = 0.20), and for different ID Selection Pool size ($\mathcal{K}$ = 400, $\mathcal{K}$ = 600, and $\mathcal{K}$ = 800), respectively. More specifically, Figure \ref{fig:r} (a) shows that as the network size increases (in this case, we set of the size of the ID Pool Selection to 800 IDs, and the resiliency of ID Selection Pool to 0.333) the failure probability ($\mathcal{P}$) delays to assume big values; this shows up the impact of the network size on the security of the network. Figure \ref{fig:r} (b) shows that when the committee resiliency gets larger the failure probability ($\mathcal{P}$) delays to assume values close to 1; this shows that a higher ID Selection Pool resiliency allows for higher security.

Figure \ref{fig:r} (c) shows the impact of the size of the ID Selection Pool on the failure probability ($ \mathcal{P}$). We observe that as the size of ID selection pool gets larger the failure probability is slow to assume values closer to 1; thus, the larger the size of the ID Selection Pool the higher the security of the network.

In addition, Figures \ref{fig:r} (d), \ref{fig:r} (e), and \ref{fig:r} (f) show the failure probability that at least one shard takeover attack ($\mathcal{P'}$) for different ID Selection Pool sizes, values of the committee resiliency, and committee sizes, respectively.

More specifically, Figure \ref{fig:r} (d) shows the impact of the size of the ID Selection Pool ($\mathcal{K}$ = 600 and $\mathcal{K}$ = 700; in this case, we set $\mathcal{N}$ to 1000, r to 0.333, and $n$ to 100) on the failure probability ($\mathcal{P'}$). We observe that as the size of ID Selection Pool gets larger the failure probability is slow to assume values closer to 1; thus, the larger the size of the ID Selection Pool the higher the security of the network. 

Figure \ref{fig:r} (e) shows the impact of the committee resiliency ($r$) on the failure probability ($\mathcal{P'}$). We observe that as the committee resiliency gets larger the failure probability is slow to assume values closer to 1; thus, the larger committee resiliency the higher the security of the network.

Figure \ref{fig:r} (f) shows the impact of the committee size ($r$) on the failure probability ($\mathcal{P'}$). We observe that as the committee size gets larger the failure probability is slow to assume large values (i.e., values closer to 1); thus, the larger the size of the committee the higher the security of the network.

Figures \ref{fig:r} (g), \ref{fig:r} (h), and \ref{fig:r} (i) show the probability of a successful attack ($\mathcal{P''}$) when varying the number of Sybil IDs ($\mathcal{M}$) from 10 to 200 by a step of 10. We observe that the failure probability ($\mathcal{P''}$) increases with the number of Sybil IDs (generated by an adversary).
This is expected, since the increase in the number of Sybil IDs leads to an increase of the chance of an adversary to compromise the security of the network.

More specifically, Figure \ref{fig:r} (g) shows the impact of the network size ($\mathcal{N}$) on the probability of a successful attack ($\mathcal{P''}$); in this case, we set $r$ to 0.25, $\mathcal{R}$ to 0.10 , $\mathcal{K}$ to 800, and $n$ to 100. We observe that as the network size gets larger the probability of a successful attack is slow to assume values closer to 1; thus, the larger size of the network the higher the security of the network.

Figure \ref{fig:r} (h) shows the impact of the committee resiliency ($r$) on the probability of a successful attack ($\mathcal{P''}$); in this case, we set $\mathcal{N}$ to 1000, $\mathcal{R}$ to 0.10, $\mathcal{K}$ to 800, and $n$ to 100. We observe that as the committee resiliency gets larger the probability of a successful attack is slow to assume values closer to 1; thus, the larger committee resiliency the higher security of the network.

Finally, Figure \ref{fig:r} (i) shows the impact of the size of the committee ($n$) on the probability of a successful attack ($\mathcal{P''}$); in this case, we set $\mathcal{N}$ to 1000, $\mathcal{R}$ to 0.10, $\mathcal{K}$ to 800, and $r$ to 0.25. We observe that as the size of the committee gets larger the probability of a successful attack is slow to assume values closer to 1; thus, the larger size of the committee the higher security of the network.

To sum up, Figure \ref{fig:r} shows that by taking advantage of PGFA we get a promising results. In particular, we identify the parameters that impact the probability of a successful attack. This means $\mathcal{N}$, $\mathcal{K}$, $\mathcal{M}$, $\mathcal{M'}$, $n$, $\lambda$, $r$, and $\mathcal{R}$.   

\begin{table*}[ht]
\centering
\caption{Computation of Years to Fail by PGFA}
\label{table:3}
\setlength{\tabcolsep}{6pt}
\begin{tabular}{|c|c|c|c|c|c|c|c|c|c|c|c|c|c|} 
\hline
$\mathcal{N}$ &
$\mathcal{K}$ &
$\mathcal{M}$ &
$\mathcal{M'}$ &
$n$ &
$\lambda$ $^{\mathrm{a}}$ &
$r$ &
$\mathcal{R}$ &
$\mathcal{P}$ &
$\mathcal{P^{'}}$ &
$\mathcal{P^{''}}$ &
$N_{s}$ &
$\mathcal{A}$ $^{\mathrm{c}}$ &
Secure \\
\hline
1000 & 
800&
200&
200&
100&
8 &
0.333&
0.20&
2.04e-06 &
1.56e-01 &
3.18e-07 &
365 &
8623.61 &
$\textcolor{green}{\checkmark } $\\
1000 & 
800&
200&
200&
100&
8 &
0.333&
0.20&
2.04e-06 &
1.56e-01 &
3.18e-07 &
185 &
17014.16 &
$\textcolor{green}{\checkmark } $\\
1400& 
800&
200&
200&
200&
8 &
0.333&
0.20&
9.25e-13 &
1.56e-01 &
1.49e-13 &
365 &
19e08 &
$\textcolor{green}{\checkmark } $ \\
1000&
800 &
200 &
200 & 
100 &
8 &
0.333 &
0.15 &
9.83e-01 &
1.56e-01 &
1.53e-01 &
365 &
1.02e-02 &
$\textcolor{red}{\times} $\\
1000 &
800&
250 & 
250 &
100 &
8 &
0.333 &
0.20 &
4.79e-01 &
9.94e-01&
4.77e-01&
365 &
3.37e-02&
$\textcolor{red}{\times} $\\
1000 &
800&
250 & 
250 & 
100 &
8 &
0.333 &
0.20 &
4.79e-01 &
9.94e-01&
4.77e-01&
185 &
1.13e-02 &
$\textcolor{red}{\times} $\\
1000 &
800 &
250 & 
125 & 
100 &
8 &
0.333 &
0.20 &
4.79e-01 &
5.69e-06 &
2.73e-06 &
185 &
1980.16 &
$\textcolor{green}{\checkmark }$\\
1000 &
800 &
250 & 
125 & 
80 &
10 &
0.333 &
0.20 &
4.79e-01 &
1.61e-04 &
7.73e-05 &
185 &
69.97 &
$\textcolor{red}{\times} $ \\
\hline
\multicolumn{10}{p{295pt}}{$^{\mathrm{a}}$ $\lambda$ = $\frac{\mathcal{K}}{n}$;
$^{\mathrm{c}}$: Years to fail, which is calculated by using the formula described in \eqref{eq:12}.}
\end{tabular}
\label{tab:Years to Fail}
\end{table*}
Table \ref{tab:Years to Fail} shows the probability of a successful attack ($\mathcal{P^{''}}$) and the average number of years to fail ($\mathcal{A}$) for
different parameters (i.e., $\mathcal{N}$, $\mathcal{K}$, $\mathcal{M}$, $\mathcal{M'}$, $n$, $\lambda$, $r$, $\mathcal{R}$, and $N_s$). We observe that when we change the values of some parameters (even small changes), the network security can be considerably impacted. For example, Table \ref{tab:Years to Fail} shows that for $\mathcal{R}$ = 0.2, the  number of years to fail is 8623.61 and for $\mathcal{R}$ = 0.15, the number of years to fail is 1.02e-02. Indeed, a network that fails 1000s years on average has a high acceptable level of security (i.e., secure) whereas a network that fails less than one year on average is not secure enough.

Now, to validate the feasibility of our approach, we perform a comparison with another existing approach called Break Consensus Protocol attack (BCP) \cite{rajab2020feasibility}. Rajab et al. \cite{rajab2020feasibility} computed (by using BCP) $\mathcal{P}$, but failed to accurately compute $\mathcal{P'}$. Thus, they failed to accurately compute $\mathcal{P''}$. BCP proposed a similar model to that of Figure \ref{fig:2}; for that reason, we compare the proposed PGFA with BCP.

\begin{figure}[ht] 
\centering
\includegraphics[width= 0.5 \textwidth]{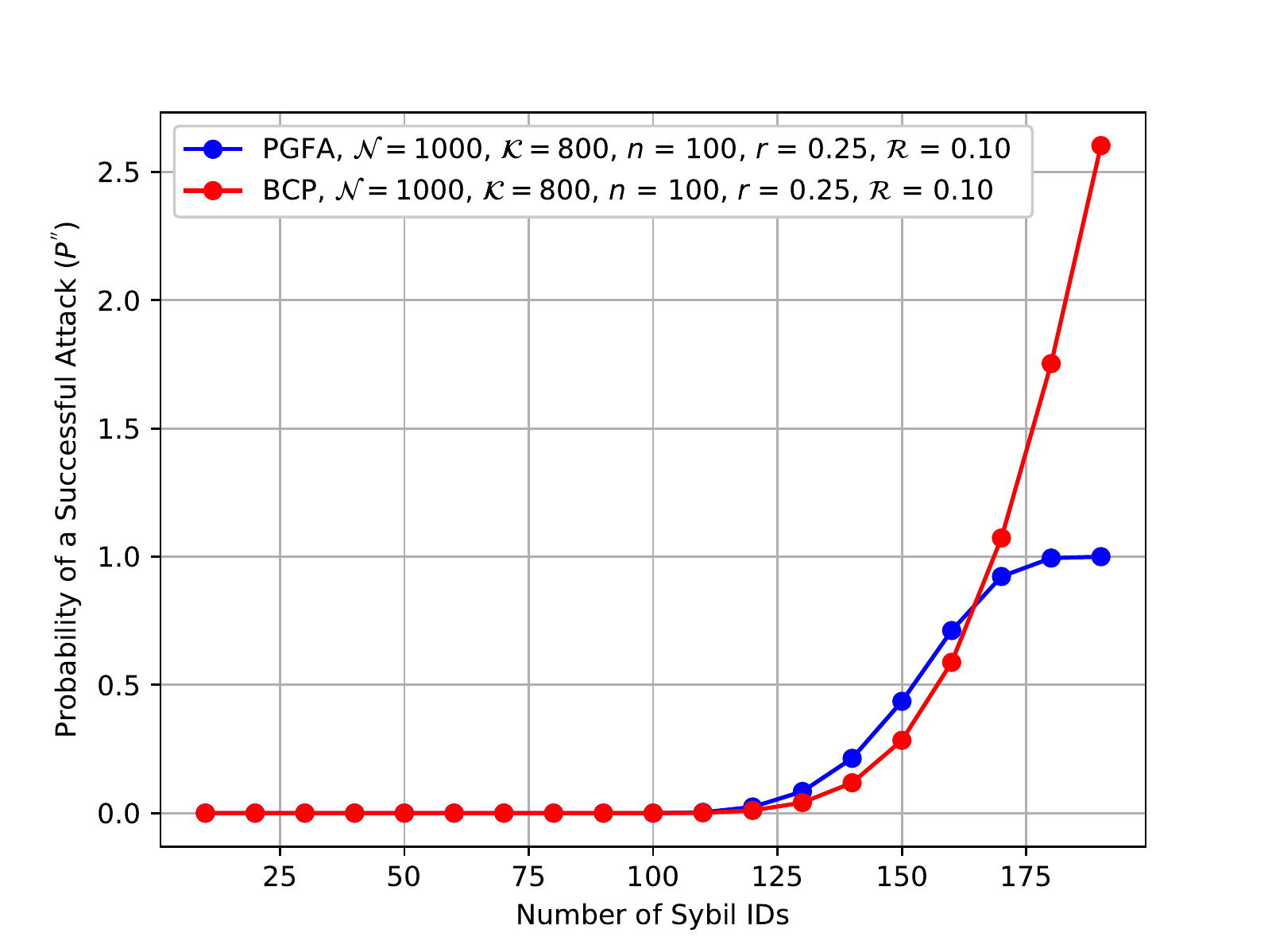} 
\caption{Comparison between PGFA and BCP}
\label{fig:comparison_BCP}
\end{figure}
Figure \ref{fig:comparison_BCP} shows the probability of a successful attack when varying the number of Sybil IDs from 10 to 200 (by a step of 10 IDs) for both PGFA and BCP approach. We observe that the probability of a successful attack computed by BCP exceeds 1. This means that BCP computes false probabilities and its formula (\textit{Theorem 1} in \cite{rajab2020feasibility}) does not corresponds to a proper probability distribution. In particular, they computed the probability that at least one shard fails by assuming that the probability of the first shard is indicative to the probability of the other shards; more specifically, they multiply the failure probability of the first shard by the number of shards to get the probability that at least one shard fails.

Note that the proposed probabilistic model can be adopted to different scenarios by adjusting some parameters (e.g. $\mathcal{K}$, $\mathcal{M^{'}}$). For instance, (1) A network with many malicious nodes where each generates numerous Sybil IDs; and (2) Honest node/nodes participate with many IDs. 

\section{Conclusion} \label{sec: Conclusion}
In this paper, we propose a tractable probabilistic approach to analyze the security of sharding-based blockchain protocols by using generating function. The proposed PGFA takes into consideration the failure probability of each committee to analyze the security instead of assuming that the failure probability for the first committee is indicative to the failure probabilities of the other committees. More specifically, we compute the failure probability that at least one committee fails. Finally, after calculating the correct failure probability (i.e., the failure probability that at least one committee fails), we propose to quantify the security of the network by computing the average number of years to failure. Furthermore, based on these probabilities, we investigate the threat of Sybil attacks. We conclude that the proposed PGFA is a promising solution to evaluate the security of existing sharding-based blockchain protocols; indeed, to the best of our knowledge, it is the most accurate and tractable approach to compute the failure probability.

\newpage

\newpage
\begin{IEEEbiography}[{\includegraphics[width=1in,height=1.25in,clip,keepaspectratio]{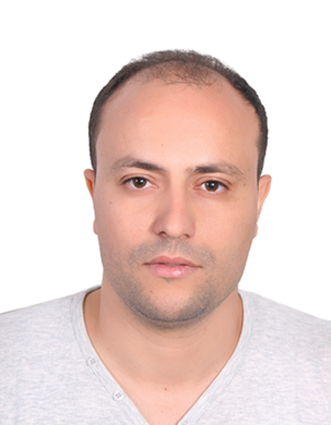}}]{Abdelatif Hafid}
received the B.Sc. degree in Mathematics and Applications from the University of Moulay Ismail, Meknes,  Morocco, and the M.Sc. degree in mathematical engineering from the University of Abdelmalek Essaâdi, Tangier, Morocco. He is currently pursuing the Ph.D. degree with the University of Moulay Ismail, Meknes, Morocco. He is also a visiting research student with the University of Montreal (UdeM), Montreal, Canada. His current research interests include Applied Probability, Statistics, and Blockchain.
\end{IEEEbiography}
\begin{IEEEbiography}[{\includegraphics[width=1in,height=1.25in,clip,keepaspectratio]{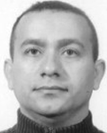}}]{Abdelhakim Senhaji Hafid}
is Full Professor at the University of Montreal. He is the founding director of Network Research Lab and Montreal Blockchain Lab. He is research fellow at CIRRELT, Montreal, Canada.  Prior to joining U. of Montreal, he spent several years, as senior research scientist, at  Bell Communications Research (Bellcore), NJ, US working in the context of major research projects on the management of next generation networks.  Dr. Hafid was also Assistant Professor at Western University (WU), Canada, Research director of Advance Communication Engineering Center (venture established by WU, Bell Canada and Bay Networks), Canada, researcher at CRIM, Canada, visiting scientist at GMD-Fokus, Germany and visiting professor at University of Evry, France. Dr. Hafid has extensive academic and industrial research experience in the area of the management and design of next generation networks. His current research interests include IoT, Fog/edge computing, blockchain, and intelligent transport systems. 
\end{IEEEbiography}
\begin{IEEEbiography}[{\includegraphics[width=1in,height=1.25in,clip,keepaspectratio]{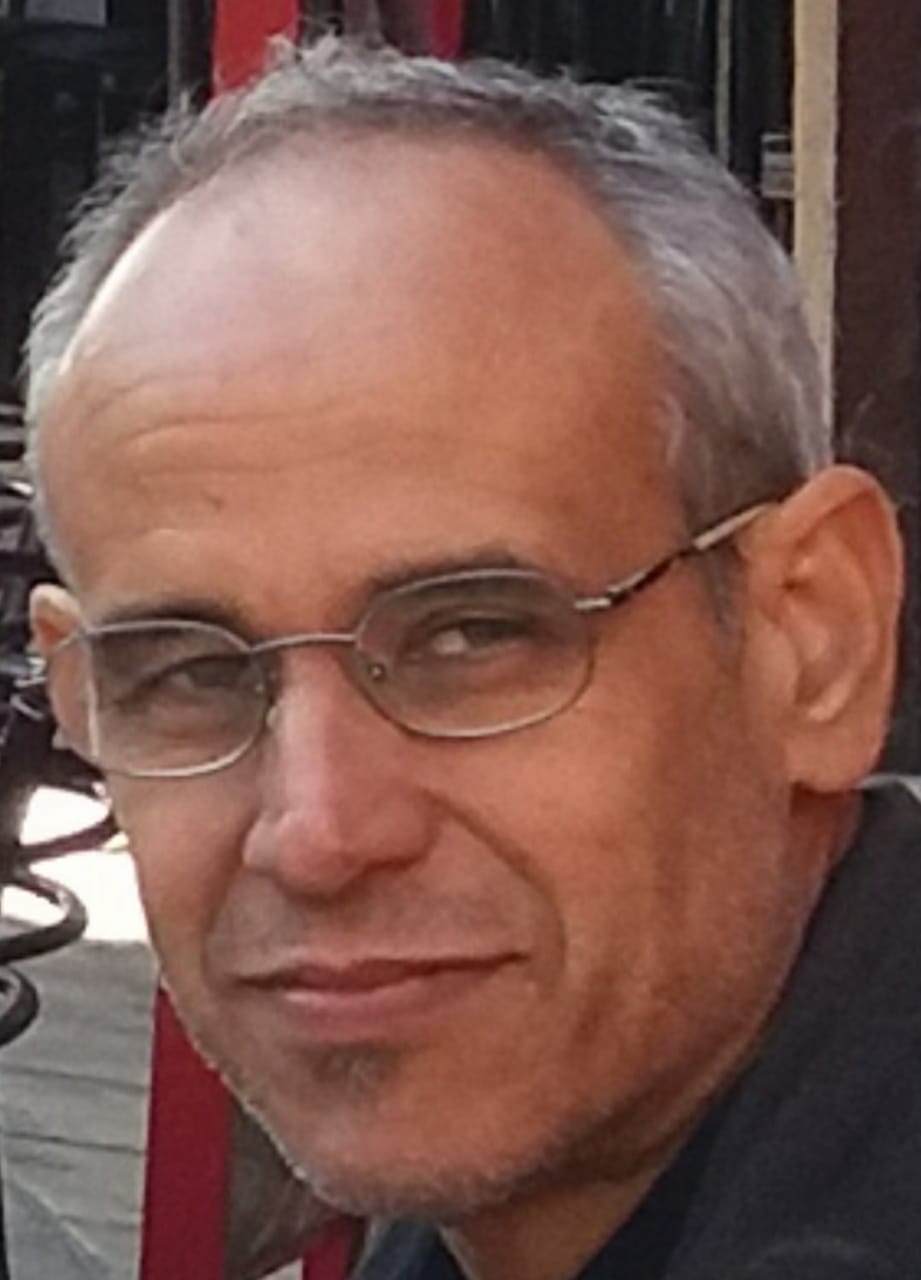}}]{Mustapha Samih}
is currently a Full Professor at the University of Moulay Ismail, Meknes,  Morocco. He received the Ph.D. degree in Fundamental and Applied Mathematics   from the University of Montpellier, France. His current research interests include Applied Probability, Statistics, and Blockchain.
\end{IEEEbiography}





\begin{thebibliography}{00}
\bibitem{bitcoin} S. Nakamoto, ``Bitcoin: A peer-to-peer electronic cash system,'' Working Paper, 2008,  [Online] Available: \textcolor{blue}{
\underline{ https://bitcoin.org/bitcoin.pdf}}

\bibitem{ethereum} G. Wood, ``Ethereum: A secure decentralised generalised transaction ledger,'' Ethereum project yellow paper, Vol. 151, pp. 1--32, 2014, [Online] Available: \textcolor{blue}{\underline{https://gavwood.com/paper.pdf}}

\bibitem{hafid2019methodology} A. Hafid, A. S. Hafid, M. Samih, ``A Methodology for a Probabilistic Security Analysis of Sharding-Based Blockchain Protocols,'' in \emph{Proceedings of the International Congress on Blockchain and Applications}, Springer, 2019, pp. 101--109. 

\bibitem{hafid2020joint} A. Hafid, A. S. Hafid and M. Samih, "A Novel Methodology-Based Joint Hypergeometric Distribution to Analyze the Security of Sharded Blockchains," in \emph{IEEE Access}, vol. 8, pp. 179389-179399, 2020, doi: 10.1109/ACCESS.2020.3027952.


\bibitem{hafid2020scaling} A. Hafid, A. S. Hafid and M. Samih, "Scaling Blockchains: A Comprehensive Survey," in \emph{IEEE Access}, vol. 8, pp. 125244-125262, 2020, doi: 10.1109/ACCESS.2020.3007251.

\bibitem{hafid2019model} A. Hafid, A. S. Hafid and M. Samih, "New Mathematical Model to Analyze Security of Sharding-Based Blockchain Protocols," in \emph{IEEE Access}, vol. 7, pp. 185447-185457, 2019, doi: 10.1109/ACCESS.2019.2961065.

\bibitem{rajab2020feasibility} T. Rajab, M. M. Manshaei, M. Jadliwala, R. Murtuza and M. A. Rahman, "On the Feasibility of Sybil Attacks in Shard-Based Permissionless Blockchains," in \emph{arXiv preprint arXiv:2002.06531}, 2020.

\bibitem{rapidchain} M. Zamani, M. Movhedi, and M. Raykova, ``Rapidchain: Scaling blockchain via full sharding,'' in \emph{Proceedings of the 2018 ACM SIGSAC Conference on Computer and Communications Security}, ACM, 2018, pp. 931--948.  

\bibitem{omniledger} E. Kokoris-Kogias, P. Jovanovic, L. Gasser, N. Gailly, E. Syta, and B. Ford, `` Omniledger: A secure, scale-out, decentralized ledger via sharding,'' in \emph{Proceedings of the 2018 IEEE Symposium on Security and Privacy (SP)}, IEEE, 2018, pp. 583--598.

\bibitem{elastico} L. Luu, V. Narayanan, C. Zheng, K.  Baweja, S. Gilbert, and P. Saxana, `` A secure sharding protocol for open blockchains,''
in \emph{Proceedings of the 2016 ACM SIGSAC Conference on Computer and Communications Security}, ACM, 2016, pp. 17--30.


\bibitem{IoT1} M. S. Ali, M. Vecchio, M. Pincheira, K. Dolui, F. Antonelli and M. H. Rehmani, "Applications of Blockchains in the Internet of Things: A Comprehensive Survey," \emph{in IEEE Communications Surveys \& Tutorials}, vol. 21, no. 2, pp. 1676-1717, Secondquarter 2019, doi: 10.1109/COMST.2018.2886932.


\bibitem{IoTsecurity} M. Khan, S. Ahmad, K. Salah, ``IoT security: Review, blockchain solutions, and open challenges," \emph{in Future Generation Computer Systems}, vol. 82, pp. 395--411, Elsevier, 2018.

\bibitem{IoT3} A. Reyna, C. Mart{\'\i}n, J. Chen, E. Soler, M. D{\'\i}az, ``On blockchain and its integration with IoT. Challenges and opportunities," \emph{in Future Generation Computer Systems }, vol. 88, pp. 173--190, Elsevier, 2018.


\bibitem{industrial} D. Mazzei, G. Baldi, F. Giacomo, M. Gualtiero, P.Gabriele, R. Antonio, R. Laura, L. Rizzello, ``A Blockchain Tokenizer for Industrial IOT trustless applications," \emph{in Future Generation Computer Systems}, vol. 105, pp. 432--445, Elsevier, 2020.

\bibitem{visa} Visa, Accessed on: Mar. 23, 2020, [Online] Available: \textcolor{blue}{\underline{https://usa.visa.com/}}

\bibitem{rajan} R. Chattamvelli, R. Shanmugam,``Generating Functions in Engineering and the Applied Sciences,'' Morgan \& Claypool, 2019. 

\bibitem{zhao2019analysis} W. Zhao, S. Jin, W. Yue,``Analysis of the Average Confirmation Time of Transactions in a Blockchain System,''
in \emph{Proceedings of the International Conference on Queueing Theory and Network Applications}, Springer, 2019, pp. 379--388. 

\bibitem{akutsu2019analysis} K. Akutsu, P. Kohei, T. Phung-Duc,``Analysis of Retrial Queues for Cognitive Wireless Networks with Sensing Time of Secondary Users,''in \emph{Proceedings of the International Conference on Queueing Theory and Network Applications}, Springer, 2019, pp. 77--91.

\bibitem{misic2019modeling} J. Misic, V. Misic, x. Chan, M. Xiaolin, S.G. Motlagh, and Z. M. Ali,``Analysis of Retrial Queues for Cognitive Wireless Networks with Sensing Time of Secondary Users,''in \emph{IEEE Transactions on Network Science and Engineering}, IEEE, 2019.

\bibitem{distinguishing} J. Wroughton and T. Cole,``Distinguishing between binomial, hypergeometric and negative binomial distributions,''in \emph{J. Statist. Educ.}, vol. 21, no. 1, 2013. 

\bibitem{lehman2010mathematics} E. Lehman, F T. Leighton, and A. R Meyer, ``Mathematics for computer science,'' \emph{MIT}, 2010.

\bibitem{poon2017plasma} J. Poon, and V. Buterin, `` Plasma: Scalable autonomous smart contracts,'' White paper, pp. 1--47, 2017, [Online] Available: \textcolor{blue}{\underline{https://plasma.io/plasma.pdf}}

\bibitem{kirti} S. B. Patel, P. Bhattacharya, S. Tanwar and N. Kumar, "KiRTi: A Blockchain-based Credit Recommender System for Financial Institutions," \emph{in IEEE Transactions on Network Science and Engineering}, doi: 10.1109/TNSE.2020.3005678.

\bibitem{HC1}  M. H. Kassab, J. DeFranco, T. Malas, P. Laplante, g. destefanis and V. V. Graciano Neto, "Exploring Research in Blockchain for Healthcare and a Roadmap for the Future," \emph{in IEEE Transactions on Emerging Topics in Computing}, doi: 10.1109/TETC.2019.2936881.
\bibitem{poon2016bitcoin} J. Poon, and T. Dryja, ``The bitcoin lightning network: Scalable off-chain instant payments,'' DRAFT Version 0.5.9.2, pp. 1--59, 2016, [Online] Available: \textcolor{blue}{ \underline{https://lightning.network/lightning-network-paper.pdf}}

\bibitem{ethereumsharding} H-W. Wang, `` Ethereum sharding: Overview and finality,'' 2017, Accessed on: Sep. 8, 2019, [Online] Available: \textcolor{blue}{\underline{https://medium.com/@icebearhww} }

\bibitem{garzik2015block} J. Garzik, ``Block size increase to 2MB,'' Bitcoin Improvement Proposal, Vol. 102, 2015.

\bibitem{network2018cheap} Raiden Network-Fast, `` cheap, scalable token transfers for Ethereum,'' 2018, [Online] Available: \textcolor{blue}{ \underline{https://raiden.network/}}

\bibitem{education} A. Alammary, S. Alhazmi, M. Almasri, S. Gillani, "Blockchain-based applications in education: A systematic review," \emph{in Applied Sciences}, vol. 9, no. 12, pp. 2400, 2019.

\bibitem{petersen2019inquiry} T K. Petersen , ``Inquiry-Based Enumerative Combinatorics: One, Two, Skip a Few... Ninety-Nine, One Hundred,'' \emph{Springer}, 2019.

\bibitem{meurer2017sympy} A. Meurer, C. P. Smith, M. Paprocki, O. {\v{C}}ert{\'\i}k, S. B. Kirpichev, M. Rocklin, A. Kumar, S. Ivanov, M. Sergiu, S. Jason K, and Others, `` SymPy: symbolic computing in Python,''
\emph{PeerJ Computer Science},  Vol. 3, pp. e103, 2017.

\bibitem{scipy} E. Bressert, `` SciPy and NumPy: An Overview for Developers,'' \emph{O’Reilly Media}, 2012.

\bibitem{ethereumsharding} V. Buterin, ``Ethereum sharding,'' 2017, [Online] Available: \textcolor{blue}{\underline{https://eth.wiki/sharding/Sharding-FAQs}}

\end{thebibliography}
\end{document}